# Multi-Label Noise Robust Collaborative Learning for Remote Sensing Image Classification

Ahmet Kerem Aksoy, *Student Member, IEEE*, Mahdyar Ravanbakhsh, *Member, IEEE*, and Begüm Demir, *Senior Member, IEEE*

*Abstract*—The development of accurate methods for multi-label classification (MLC) of remote sensing (RS) images is one of the most important research topics in RS. The MLC methods based on convolutional neural networks (CNNs) have shown strong performance gains in RS. However, they usually require a high number of reliable training images annotated with multiple land-cover class labels. Collecting such data is time-consuming and costly. To address this problem, the publicly available thematic products, which can include noisy labels, can be used to annotate RS images with zero-labeling cost. However, multi-label noise (which can be associated with wrong and missing label annotations) can distort the learning process of the MLC methods. To address this problem, we propose a novel multi-label noise robust collaborative learning (RCML) method to alleviate the negative effects of multi-label noise during the training phase of a CNN model. RCML identifies, ranks, and excludes noisy multi-labels in RS images based on three main modules: 1) the discrepancy module; 2) the group lasso module; and 3) the swap module. The discrepancy module ensures that the two networks learn diverse features, while producing the same predictions. The task of the group lasso module is to detect the potentially noisy labels assigned to multi-labeled training images, while the swap module is devoted to exchange the ranking information between two networks. Unlike the existing methods that make assumptions about noise distribution, our proposed RCML does not make any prior assumption about the type of noise in the training set. The experiments conducted on two multi-label RS image archives confirm the robustness of the proposed RCML under extreme multi-label noise rates. Our code is publicly available at: http://www.noisy-labels-in-rs.org.

*Index Terms*—Collaborative learning, deep learning (DL), multi-label image classification, multi-label noise, remote sensing (RS).

## I. INTRODUCTION

REMOTE sensing (RS) images acquired by satellite-borne and airborne sensors are a rich source of information for monitoring the Earth's surface, e.g., urban area studies, forestry applications, and crop monitoring [1]. As a result of

Manuscript received 22 December 2020; revised 16 December 2021, 18 March 2022, and 22 June 2022; accepted 17 September 2022. This work was supported in part by the European Research Council (ERC) through the ERC-2017-STG-BigEarth Project under Grant 759764, in part by the German Ministry for Education and Research through the Berlin Institute for the Foundations of Learning and Data (BIFOLD) under Grant 01IS18025A, and in part by the TreeSatAI Project under Grant 01IS20014A. *(Corresponding author: Begüm Demir.)*

The authors are with the Faculty of Electrical Engineering and Computer Science, Technische Universität Berlin, 10623 Berlin, Germany (e-mail: a.aksoy@tu-berlin.de; ravanbakhsh@tu-berlin.de; demir@tu-berlin.de).

Color versions of one or more figures in this article are available at https://doi.org/10.1109/TNNLS.2022.3209992.

Digital Object Identifier 10.1109/TNNLS.2022.3209992

recent advances in RS technology, huge amounts of RS images have been acquired and stored in massive archives, from which mining of useful information is an important and challenging issue. Specifically, the development of accurate multi-label RS image scene classification methods that automatically assign multiple land-cover class labels (i.e., multi-labels) to each RS image scene in an archive is a growing research interest in RS. In recent years, DL approaches have attracted great attention also in the multi-label classification (MLC) of RS images due to their high capability to describe the complex spatial and spectral content of RS images. As an example, in [2] a convolutional neural network (CNN) that includes a softmax function as the activation of the last CNN layer is presented. In [3], a radial basis function neural network with an MLC layer is proposed. In [4], an attention-based long short-term memory (LSTM) network is used to sequentially predict classes one after another. An encoder–decoder neural network that includes: 1) a squeeze excitation layer (which characterizes channelwise interdependencies of image feature maps) and 2) an adaptive spatial attention mechanism (which models the informative image regions) is proposed in [5]. A multiattention-driven approach that contains: 1) spatial-resolution-specific CNNs in a branchwise architecture and 2) a bidirectional LSTM network is presented in [6]. A study to analyze and compare different DL loss functions in the framework of MLC of RS images is presented in [7].

Most of the above mentioned DL-based approaches for the MLC of RS images require a sufficient number of high-quality (i.e., reliable) training images annotated with multi-labels. This is crucial for accurately characterizing complex content of images with discriminative and descriptive features and thus for achieving accurate multi-label predictions. However, collection of multi-labeled images is time-consuming, complex, and costly in operational scenarios and can significantly affect the final accuracy of the MLC methods [8]. To address this problem, a common approach is to use the DL models pretrained on publicly available datasets in the computer vision (CV) community (e.g., ImageNet [9]). Then, the final layers of the pretrained models are fine-tuned by considering a small set of multi-labeled RS images for the target classification task. This approach is not optimal for RS images due to different image characteristics in CV and RS (e.g., Sentinel-2 multispectral images have 13 spectral bands associated with varying and lower spatial resolutions compared with the images in CV). In addition, the semantic content (and thus the considered semantic classes) present in RS images is significantly different from that of CV images (see Fig. 1). Thus, the DL models trained from scratch on a large RS training set annotated with multi-labels are required.

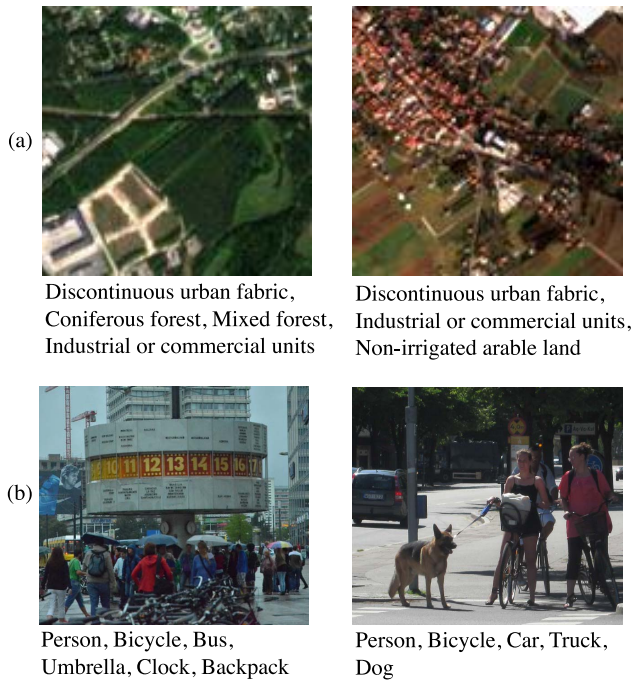

Fig. 1. Example of training images annotated with multi-labels in (a) RS [8] and (b) CV [15].

An effective approach for constructing a large training set with zero-annotation effort is to exploit the publicly available thematic products [e.g., the Corine Land Cover (CLC) map [10], the GLC2000, and the GlobCover] in RS as labeling sources [8], [11]. As an example, Sumbul et al. [8] develop the BigEarthNet benchmark archive made up of Sentinel-2 multispectral images annotated using the CLC map to drive the DL studies in RS MLC. Constructing such large RS training sets with zero labeling cost is highly valuable. However, the set of land-cover class labels available in a given area through the thematic products can be incomplete or wrong (i.e., noisy). As an example, according to the validation report of CLC, the accuracy is around 85% [12]. Using training images with noisy labels may result in uncertainty in the MLC model and thus may lead to reduced performance on multi-label prediction. Accordingly, methods that reduce the negative impact of noisy annotations are needed in the framework of RS MLC. When the training images are associated with multi-labels, two types of noise can exist for a given RS image.

1) Noise associated with missing labels: This type of noise appears when a land-cover class label is not assigned to an image while that class is present in the image (i.e., the class label is missed from the label set of that image) [13].

2) Noise associated with wrong labels: This type of noise appears when a land-cover class label is assigned to an image although that class is not present in the given image (i.e., the label is wrongly assigned in the label set of that image) [14].

The number of missing or wrong class labels can vary depending on the labeling source. Two types of noise can also simultaneously appear associated with different spatial areas of a given RS image (e.g., a land-cover class label can be missed, while another land-cover class label is wrongly assigned in different spatial portions of the image). Since the DL models can easily overfit to noisy labeled data [16], [17], dealing with label noise can significantly improve the MLC performance. Recently, a couple of studies in RS are presented to learn from noisy labels in RS MLC. As an example, in [18], a semantic segmentation method that identifies label noise is presented to generate accurate land-cover maps by classifying RS images. This is achieved by simply evaluating the loss values since the noisy image labels are associated with the highest values of the loss. However, this method can only identify the wrong label noise and ignores the missing label noise problem. Hua et al. [19] propose a regularization method to improve the MLC performance in RS under label noise. Regularization is defined on the basis of a label correlation matrix, which is constructed by measuring the distances between the corresponding word vectors in a text embedding space. Construction of a reliable label correlation matrix for different RS applications is a complex task due to the difficulty in collecting text descriptions of class labels for properly modeling the correlation between all possible combinations of classes present in RS images. The performance of these two methods depends on accurate estimation of noise distribution in the considered data. Thus, there is a need to make prior assumptions about the noise type, which restricts the applicability and generalization capability of the methods for different MLC applications with different noise distributions. This is a critical limitation for complex RS MLC problems. Unlike RS, in the CV community, the development of noise robust DL models is much more extended and widely studied (see Section II-A for the literature survey). However, most of the existing methods assume that each image is annotated by a single label associated with the most significant content of the considered image [20], [21]. Adapting single label noise tolerance methods for multi-labeled images is a challenging task due to the complexity of modeling the above-mentioned two types of noise in multi-labeled images. This becomes more critical when the number of land-cover classes (and thus class combinations) increases.

To address this problem, we propose a novel multi-label noise robust collaborative learning (RCML) method. Unlike the existing methods in the literature, the proposed method identifies, ranks, and excludes samples with noisy multi-labels in RS images without making any prior assumption about the type of noise in the training set. To this end, the proposed RCML method contains three modules: 1) discrepancy module; 2) group lasso module; and 3) swap module. The discrepancy module aims at forcing the two networks to learn diverse features, while achieving consistent predictions. To automatically identify the noise type, we propose a group lasso module that computes a samplewise ranking loss. The swap module exchanges the ranking information between the two collaborative networks and excludes extremely noisy labeled images from backpropagation. Our proposed collaborative learning method has been briefly introduced in [22], which was only applicable for training sets with a high noise rate. This article extends the previous work by addressing the limitations of the earlier method and includes a detailed description of each step with an extended experimental analysis. The main contributions of this work are summarized as follows.

1) To the best of our knowledge, RCML is the first method that automatically identifies the two types of noise in multi-labeled RS images without any prior assumption about noise distribution.

2) RCML can learn from highly noisy training sets by identifying noisy samples and excluding them from the training process.

3) We adapt a collaborative learning approach to be suitable for multi-labeled images based on ranking training images according to their identified label noise and exchanging this ranking information.

4) The proposed RCML is an architecture-independent method, which can be integrated within different classification approaches.

The rest of this article is organized as follows. In Section II, we survey the DL-based methods presented in the CV community for learning from noisy single-labeled and multi-labeled training images. In Section III, we introduce the proposed RCML method in detail. Section IV describes the considered datasets and the experimental setup, while Section V represents the experimental results. Finally, Section VI concludes this article.

## II. Related Work

In this section, we review the methods that are robust to label noise and are presented in the CV literature in the context of image scene classification. We categorize the considered methods into two groups as: 1) methods robust to noise on single-labeled images and 2) methods robust to noise on multi-labeled images.

### A. Methods Robust to Noise on Single-Labeled Images

When an image is annotated with a single label, there is only one type of noise associated with the wrong class label assignment. The methods in this category aim to model the noise distribution present in the data considering only this type of noise to reduce the influence of noisy labels on the result. Patrini et al. [23] propose a loss correction method by estimating the noise transition matrix of data. This method aims at fixing the class-dependent label noise in the data, given the probability of each class being wrongly assigned into another. In [24], the correct labels are considered latent variables in the presence of noisy labels, and the expectation–maximization [25] algorithm is applied to iteratively calculate the correct labels and the network parameters. This scheme is also extended to scenarios, where noisy labels are dependent on the features, by modeling noise via a softmax layer that connects correct labels to noisy labels. In [26], it is shown that using the common overfitting prevention techniques (e.g., regularization and dropout) can be partially effective when dealing with label noise. However, in these cases, label noise may still affect the performance of the classifier, resulting in an accuracy drop. Moreover, since noise injection is already an overfitting prevention technique that attempts to improve the generalization performance [27], there is a need to apply appropriate regularization to prevent underfitting of the classifier. In [28], an early learning regularization (ELR) method is proposed to use the memorization effects of the DL models. The ELR introduces a regularization term that aims to identify the more important parameters for learning the clean labels before early stopping, and then deactivating the unimportant parameters to prevent model fitting on noisy labels. Sukhbaatar et al. [29] explore the performance of CNNs trained on noisy labels. CNNs are modified by including an additional fully connected layer at the end of their network architecture to adapt the predictions to the noisy label distribution of the data. In [30], the training set is partitioned into multiple subsets to train multiple classifiers. If all the classifiers agree on a label from the original training set, the label is updated with the agreed label prediction. This process is repeated over several stages, which gradually improve the overall performance. Nonetheless, this process depends on the performance of the individual classifiers, and it can be time-demanding when large and complex training sets are used. Dehghani et al. [31] use a student model trained on noisy data along with a teacher model that exploits the noise structure extracted by the student model. However, this approach needs training samples with both clean and noisy labels, which are not always available.

In [32], a self-adaptive training (SAT) algorithm is proposed to dynamically correct noisy training samples. To this end, the SAT uses an exponential moving average of the model predictions to improve the generalization of deep networks under label noise. Similarly, in [33], a self-error-correcting CNN is proposed that can operate on highly noisy data. The network swaps potential noisy labels with the most probable prediction of the network, while simultaneously optimizing the model parameters. In [34], a simple strategy to separate the noisy labels from clean labels is introduced. This strategy uses a hard-sample mining technique based on the focal loss (FL) [35] to distinguish clean samples from noisy ones during the early stage of training. Failing to distinguish hard samples from noisy ones is a common problem, which may result in an undesired situation where the model swaps labels by mistake. To address this issue, Wu et al. [36] use semantic bootstrapping to detect noisy samples and remove them from training. Removing as few samples as possible is an important aspect of the noisy sample removal process to avoid unnecessary information loss. In [37], a deep neural network (DNN) is trained in an unsupervised fashion over a noisy training set by zero-weighting the noisy labels while keeping the samples in the training process. This prevents losing training samples, and thus allows DNNs to learn from more samples.

The idea of curriculum learning using multiple networks is exploited in MentorNet [38]. MentorNet imposes a data-driven sample weighting curriculum learning on the student network to choose clean samples from the data. Han et al. [39] propose a learning paradigm called co-training. Under co-training, two networks are trained simultaneously, choosing batches that include only clean training samples to feed each other. Each network back-propagates the mini-batch that is chosen by its peer network to update itself. Han et al. demonstrate that co-training is an effective method when dealing with label noise. In [40], a co-training method is proposed with an additional discrepancy measurement to enforce two networks diverge. The two peer networks that learn different features of the same probability distribution are used. Then, both the networks select clean samples for each other based on their predictions to improve the performances mutually. Ren et al. [41] propose a meta-learning algorithm, in which a weighting factor is assigned to each training sample based on its gradient direction. Although this method achieves impressive results, it needs a small set of clean training samples, which may not always be present. One of the advantages of the co-training methods is that they can be decoupled from the training process of any specific classifier. Such decoupled label noise estimation process allows the co-training approaches to be used and tested with different network architectures. However, the success of the above-mentioned works mostly depends on the accurate estimation of noise distribution in data, which restricts the applicability and generalization capability of these methods. Also, most of the above-mentioned methods are designed to address only one type of noise (i.e., wrong class label assignment) and are not directly applicable for the MLC problems in RS.

### B. Methods Robust to Noise on Multi-Labeled Images

There are only few works presented in the literature to address the multi-label noise problems. As an example, Ghosh et al. [42] study different loss functions such as categorical cross-entropy (CCE), mean square error (MSE), and mean absolute error (MAE) for noise robust MLC and argue that MAE is more robust to label noise compared with the other loss functions. In [43], it is stated that MAE shows poor performance when complex training sets are considered. To address this issue, a set of robust loss functions that combine CCE and MAE is proposed as noise robust alternatives to CCE [42]. Meta-learning and ensemble methods are other approaches proposed to address the label noise in MLC problems. Li et al. [44] aim to find noise-tolerant model parameters using a teacher model along with a student model to make accurate predictions by optimizing a meta-objective, which encourages the student model to give consistent results with the teacher model after introducing synthetic labels.

Bucak et al. [45] present a ranking-based multi-label learning method that exploits the group lasso to enhance the accuracy with missing class labels. This method initially computes two error values: 1) error associated with predicted classes in the multi-label set and 2) error for the unpredicted ones within the same multi-label set for each sample. The two errors are then combined to define the ranking error for each unpredicted class. This error value indicates the possible missing class for the related sample. Finally, all the ranking errors associated with that sample are summed to define a final ranking error. The group lasso is introduced in [46] as an extension to the regular lasso, grouping a set of variables together for accurate prediction in regression. It is effective, because it groups variables together to be included or excluded completely, as opposed to regular lasso, which only selects variables individually. The group lasso is used to regularize the network in [45], giving empirically robust results against missing class labels. Durand et al. [14] propose a modified binary cross-entropy loss, which reduces the negative effects of missing class labels during training. Jain et al. [13] address the problem of missing class labels by defining a novel loss function called propensity scored loss (PSL). PSL can prioritize the most relevant labels over the many irrelevant ones and provide an unbiased estimation of true labels without omitting the missing labels entirely. Xie and Huang [47] present an approach for partial multi-label learning with noisy label identification that simultaneously recovers the ground-truth labeling information and identifies the noisy labels. This is achieved under the supervision of the observed noise-corrupted label matrix. In [48], it is assumed that annotators provide only one relevant label for each image. Then, the existing multi-label losses are extended to this setting, and variants are introduced that constrain the number of expected class labels present in the image (i.e., positive labels) during training.

As mentioned before, two types of noise (which are associated with wrong and missing class labels) may exist when training images annotated with multi-labels are considered. However, all the aforementioned methods are designed to overcome only one type of label noise (either missing or wrong class labels) in the training sets and are not capable of identifying the two noise types simultaneously. This is an important limitation in MLC. In this article to address this issue, we propose a novel collaborative learning method that aims at training a DL model robust to the two types of label noise for an accurate MLC of RS images.

### III. PROPOSED METHOD

The proposed RCML method consists of three main modules: 1) discrepancy module; 2) group lasso module; and 3) swap module. The discrepancy module is devoted to allow the two networks that are used collaboratively in the method to learn different feature sets, while ensuring consistent predictions. The group lasso module aims at identifying potential noisy class labels by computing a samplewise ranking loss. The swap module aims at exchanging the ranking information between the networks using the ranking loss functions and excluding the detected noisy samples in the final loss calculation. RCML follows the principle of a collaborative framework called co-training to exchange ranking information between networks and to select samples associated with small loss values from the training set. Co-training is a semisupervised learning technique that was first proposed in [49] to overcome labeled data insufficiency. In [50], the co-training framework is extended to apply deep networks to the task of semisupervised image recognition. In detail, deep co-training that trains multiple DNNs to be of different views and exploits adversarial examples to encourage view difference to avoid the networks from collapsing into each other is introduced.

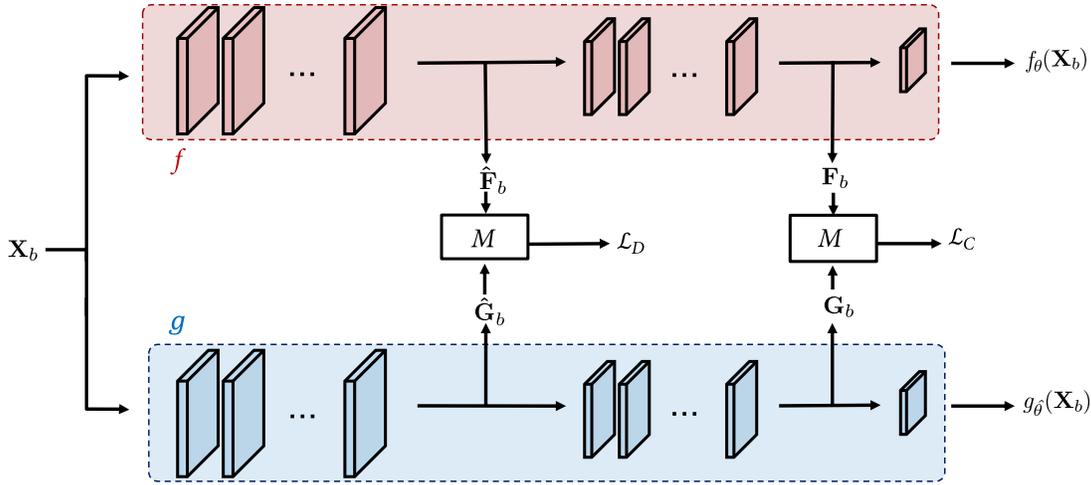

Fig. 2. Block diagram of the proposed discrepancy module. $\mathbf{X}_b$ denotes the set of training images in the given mini-batch $b$. $f$ and $g$ represent the networks. $\hat{\mathbf{F}}_\mathbf{b}$ and $\hat{\mathbf{G}}_\mathbf{b}$ stand for the intermediate logits of the networks. $\mathbf{F}_\mathbf{b}$ and $\mathbf{G}_\mathbf{b}$ stand for the logits of the last layers. $M$ is the discrepancy module. $\mathcal{L}_D$ is the disparity loss, and $\mathcal{L}_C$ is the consistency loss.

Inspired by [40], we use two CNNs with the same architectures, which learn independent set of features, while attaining the same class distribution. Therefore, it becomes easier for the networks to find their potential faults. This improves the ability of selecting training images with clean labels immensely, since two networks are forced to learn different features and correct each other by exchanging their loss information. It is worth noting that our proposed collaborative model is architecture-independent, since it does not rely on any specific network architecture. The RCML modules are applicable in the framework of any classification algorithm to detect the potentially noisy labels assigned to the training images with multi-labels.

### A. Problem Formulation

We consider a multi-label image classification problem with a training set $\mathcal{D} = \{(x_i, y_i)\}_{i=1}^D$, where $x_i$ denotes the $i^{th}$ training image. Each training image is associated with one or more classes from a set of labels $\{l_1, l_2, \ldots, l_V\}$ where $V$ is the total number of classes. $y_i = [y_i^v]_{v=1}^V \in \{0, 1\}^V$ is a binary vector, where $y_i^v$ indicates the presence or absence of the $v$th label for the image $\mathbf{x}_i$. We assume that the labels in $y_i$ can be noisy (e.g., wrong or missing). To reduce the negative effect of a noisy label, we propose a novel RCML method, based on two collaborative CNNs. In detail, we use two CNNs with identical architectures that are represented as $f$ and $g$ with parameters $\theta$ and $\hat{\theta}$, respectively. Loss functions $\mathcal{L}_{\text{clc}}^f$ and $\mathcal{L}_{\text{clc}}^g$ are the classification loss functions for networks $f$ and $g$, respectively. For each network, the binary cross entropy (BCE) is chosen for the classification loss function as suggested in [8]. The considered collaborative networks aim at: 1) learning diverse features, while producing consistence class predictions and 2) being capable of correcting errors of each other during training through information exchange. To this end, the proposed RCML contains three main modules: 1) discrepancy module; 2) group lasso module; and 3) swap module. Each module is described in detail in the following subsections.

### B. Discrepancy Module

In a collaborative learning framework, the networks must learn independent features while predicting the same class distribution. The networks are required to be capable of fixing mistakes of each other in the training process by exchanging information. If they do not learn diverse features on the same mini-batch, they cannot enhance predictions of each other mutually. To achieve the desired diversity, we propose a discrepancy module embedded in our method, which includes two loss functions: 1) the disparity loss ($\mathcal{L}_D$) and 2) the consistency loss ($\mathcal{L}_C$). The disparity loss function ensures that the networks learn distinct features, while the consistency loss function ensures that the two networks produce similar predictions. Therefore, for each mini-batch $b$, the disparity loss is calculated between the chosen layers of the networks, and the consistency loss is calculated at the end of the networks (See Fig. 2). The disparity loss $\mathcal{L}_D$ is defined as

$$\mathcal{L}_D = M(\hat{\mathbf{F}}_b, \hat{\mathbf{G}}_b)$$
$$\hat{\mathbf{G}}_b = g_{\hat{\theta}(1:c)}(\mathbf{X}_b), \hat{\mathbf{F}}_b = f_{\theta(1:c)}(\mathbf{X}_b) \quad (1)$$

where $M$ is the discrepancy module. $\hat{\mathbf{F}}_b$ and $\hat{\mathbf{G}}_b$ represent the logits of the layers before the module. The parameters of the networks $f$ and $g$ up to the layer $c$ are denoted as $\theta(1:c)$ and $\hat{\theta}(1:c)$, respectively. $c$ is the layer that the discrepancy module for the disparity loss is inserted. The set of training images in the given mini-batch $b$ is denoted as $\mathbf{X}_b$. The consistency loss $\mathcal{L}_C$ is defined as

$$\mathcal{L}_C = M(\mathbf{F}_b, \mathbf{G}_b) \quad (2)$$

where $\mathbf{F}_b$ and $\mathbf{G}_b$ denote the logits of the last layer of the networks. The discrepancy module is a statistical distance function, which measures the difference between two probability distributions. In general, for the discrepancy module $M$, any statistical distance function can be used (e.g., maximum mean discrepancy (MMD) [51] and Wasserstein metric [52]). We select the MMD algorithm to be used in the discrepancy module due to its success to disentangle probability distributions [40]. The MMD is defined as the distance between the mean embeddings of the distributions in reproducing Kernel Hilbert space (RKHS) [53]. In detail, MMD for two distributions $P$ and $Q$ is defined as

$$\text{MMD}(P, Q) = \|\boldsymbol{\mu}_P - \boldsymbol{\mu}_Q\|_H \tag{3}$$

where $\boldsymbol{\mu}_P$ and $\boldsymbol{\mu}_Q$ denote the mean values of the distributions $P$ and $Q$, respectively. $H$ denotes the RKHS, and $\|\ \|_H$ represents the $L_1$-norm. An empirical estimation of MMD between $P$ and $Q$ can be denoted as

$$\text{MMD}(P, Q) = \frac{1}{m^2}\left[\sum_{i=1}^{m}\sum_{t=1}^{m}k(\mathbf{s}_j^P, \mathbf{s}_t^P) - 2\sum_{i=j}^{m}\sum_{t=1}^{m}k(\mathbf{s}_j^P, \mathbf{s}_t^Q)\right.$$
$$\left. + \sum_{j=1}^{m}\sum_{t=1}^{m}k(\mathbf{s}_i^Q, \mathbf{s}_t^Q)\right] \tag{4}$$

where $\mathbf{s}_j^P$ and $\mathbf{s}_j^Q$ are the $j$th samples from respective distributions. $k$ is the Gaussian radial basis function kernel [54].

### C. Group Lasso Module

The BCE loss function is a widely used objective function in multi-label learning. However, while training set contains label noise, the networks may be biased toward noise in the training set and perform poorly. Thus, an additional mechanism is necessary to avoid the model misguided by noisy training sets. The additional mechanism can have many forms, such as regularization, noisy labeled image exclusion, or noise correction. Inspired by [45], we introduce a ranking error function capable of dealing with two types of noise in a multi-label training set (missing label and wrong label) without considering prior assumption. To this end, the ranking error function for missing class labels proposed in [45] is extended to identify wrong class label assignments as well. In addition, we do not use our ranking error function as a regularizer. Instead, we use it to detect noisy labels within a mini-batch and exclude them from backpropagation. Excluding the training images with noisy labels in backpropagation prevents overfitting to noisy training samples. The motivation behind using group lasso is to identify potential noisy labels in the training set, given the opportunity to learn from those samples. Furthermore, it provides information about label noise type. Let $\mathcal{E}_{l,\hat{l}}^f$ denote the ranking error function for network $f$

$$\mathcal{E}_{l,\hat{l}}^f(\mathbf{x}_i) = \max\left(0, [2(f_{\hat{l}}(\mathbf{x}_i) - f_l(\mathbf{x}_i)) + 1]\right) \tag{5}$$

where $l \in L$ and $\hat{l} \in \hat{L}$ denote the assigned and unassigned labels to the image $\mathbf{x}_i$, respectively. $f_l(\mathbf{x}_i)$ and $f_{\hat{l}}(\mathbf{x}_i)$ denote the prediction probabilities from network $f$ for the classes $l$ and $\hat{l}$, respectively. The ranking error function gives a measure of potential noise in a class combination. In the case of correct prediction, the ranking error is equal to 0; otherwise, the ranking error function returns a positive value indicating a label noise. The values from ranking error functions are gathered by two loss terms using the group lasso to identify the potential label noise. The ranking loss for network $f$ is defined as

$$\text{Lasso}_f(\mathbf{x}_i) = \alpha\sum_{\hat{l} \in \hat{L}}\sqrt{\sum_{l \in L}\mathcal{E}_{l,\hat{l}}^2(\mathbf{x}_i)} + \beta\sum_{l \in L}\sqrt{\sum_{\hat{l} \in \hat{L}}\mathcal{E}_{l,\hat{l}}^2(\mathbf{x}_i)} \tag{6}$$

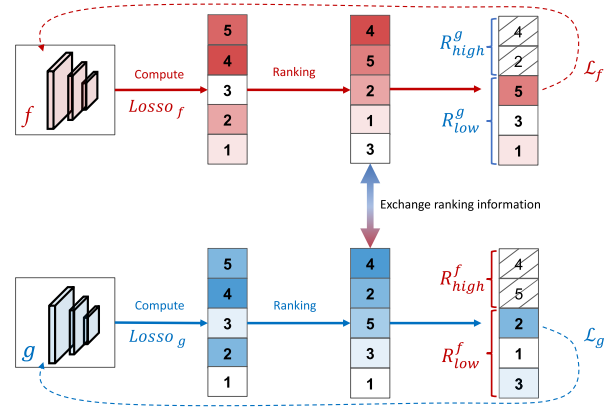

Fig. 3. Qualitative example to describe the swap module. The two networks $f$ and $g$ exchange the ranking information. Network $f$ updates its weights by calculating the final loss $\mathcal{L}_f$, using the samples with smaller ranking loss values $R_{\text{low}}^g$ that are identified by $g$. Similarly, network $g$ updates its parameters by calculating the final loss $\mathcal{L}_g$ with respect to $R_{\text{low}}^f$ identified by $f$ from backpropagation. For this qualitative example, network $f$ calculates the final loss as $\mathcal{L}_f = \lambda_1\mathcal{L}_{\text{clc}}^f(\{x_1, x_3, x_5\}) + \lambda_2\mathcal{L}_C - \lambda_3\mathcal{L}_D$ to update its parameters.

where the first loss term calculates an aggregated loss based on missing class labels, while the second term calculates an aggregated loss for wrong class label assignments. This approach allows our method to rank the training samples associated with noisy labels according to their estimated noise rate and noise type by adjusting the importance factors ($\alpha$ and $\beta$) of the loss terms. Similarly, for network $g$, the ranking loss $\text{Lasso}_g(\mathbf{x}_i)$ is computed analogous to $\text{Lasso}_f(\mathbf{x}_i)$. Then, the calculated ranking losses obtained from each network are sent to the swap module to identify the training samples associated with noisy labels.

### D. Swap Module

As shown in Fig. 3, the swap module is injected between the two collaborative networks and aims at exchanging the ranking information obtained from the group lasso module. In detail, the swap module uses the ranking losses calculated by (6) and sorts them by the ascending order. Then, the swap module splits the obtained ranking losses from each collaborative network into two sets: 1) samples associated with highest ranking loss values ($R_{\text{high}}^f$ and $R_{\text{high}}^g$, for networks $f$ and $g$, respectively) and 2) samples associated with smaller ranking loss values ($R_{\text{low}}^f$ and $R_{\text{low}}^g$, for $f$ and $g$, respectively). Then the ranking information of samples associated with the highest and lowest loss values are exchanged between the two networks. To calculate the final loss for each network, the classification loss is only computed for the samples associated with lowest ranking loss values (i.e., identified samples with clean labels). In detail, network $f$ calculates the final loss $\mathcal{L}_f$ using $R_{\text{low}}^g$ identified by the network $g$, and vice versa. A visualization of the process is illustrated in Fig. 3. The final losses for two networks $f$ and $g$ are defined as

$$\mathcal{L}_f = \lambda_1\mathcal{L}_{\text{clc}}^f(x_r^g) + \lambda_2\mathcal{L}_C - \lambda_3\mathcal{L}_D \quad \forall x_r^g \in R_{\text{low}}^g$$
$$\mathcal{L}_g = \lambda_1\mathcal{L}_{\text{clc}}^g(x_r^f) + \lambda_2\mathcal{L}_C - \lambda_3\mathcal{L}_D \quad \forall x_r^f \in R_{\text{low}}^f \tag{7}$$

where $\lambda_1$ is a weight value for the BCE loss calculated over the identified samples associated with the lowest ranking loss values. $\lambda_2$ and $\lambda_3$ represent the weights for $\mathcal{L}_C$ and $\mathcal{L}_D$, respectively. The number of samples associated with smaller ranking loss (i.e., $R_{\text{low}}^g$ and $R_{\text{low}}^f$) is defined by an adaptive swap rate $\gamma$. The swap rate indicates the contribution of the identified clean samples in the final loss value. It worth noting that the disparity loss $\mathcal{L}_D$ forces two networks learn distinct features, and therefore, it is maximized. The consistency loss $\mathcal{L}_C$ is minimized to ensure that the networks produce similar predictions. Through an end-to-end training of the two collaborative networks, the parameters of the networks $f$ and $g$ are learned by minimizing the $\mathcal{L}_f$ and $\mathcal{L}_g$ losses, respectively. After the training phase, both the networks are used to classify each image in the archive.

## IV. Dataset Description and Experimental Setup

### A. Dataset Description

We conducted experiments on two different benchmark RS datasets. The first dataset is the Ireland subset of the BigEarthNet (denoted as IR-BigEarthNet) benchmark archive [8], which consists of 15 894 Sentinel-2 multispectral images acquired between June 2017 and May 2018 over Ireland. Each image was annotated with multiple land-cover classes provided by 2018 CLC Map inventory. Sentinel-2 images contain 13 spectral bands with varying spatial resolutions. Each image in IR-BigEarthNet is a section of: 1) $120 \times 120$ pixels for the bands that have a spatial resolution of 10 m; 2) $60 \times 60$ pixels for the bands that have a spatial resolution of 20 m; and 3) $20 \times 20$ pixels for the bands that have a spatial resolution of 60 m. In the experiments, we used the bands with 10- and 20-m spatial resolutions (and thus ten bands per image is used in total), excluding the two 60-m bands due to their very low spatial resolution and small pixel size. The cubic interpolation was applied to 20-m bands of each image to have the same pixel sizes associated with each band. In the experiments, we exploited the 19 land-cover class nomenclature proposed in [55] for BigEarthNet and eliminated seven classes which are represented with a significantly small number of images in the dataset, leading to 12 classes in total. The number of labels associated with each image varies between 1 and 7, while 97,4% of images contain less than five labels. IR-BigEarthNet was divided into a validation set of 3839 images, a test set of 3856 images, and a training set of 8192 images. In the experiments, the images that are fully covered by seasonal snow, cloud, and cloud shadow were not used as suggested in [8]. An example of images from IR-BigEarthNet together with their multi-labels is given in Fig. 4. Table I shows the number of training, validation, and test samples associated with each considered class in IR-BigEarthNet.

The second dataset is the UCMerced Land Use (denoted as UCMerced) archive that consists of 2100 images selected from aerial orthoimagery and downloaded from the United States Geological Survey (USGS) National Map of the following US regions: Birmingham, Boston, Buffalo, Columbus, Dallas, Harrisburg, Houston, Jacksonville, Las Vegas, Los Angeles, Miami, Napa, New York, Reno, San Diego, Santa Barbara,

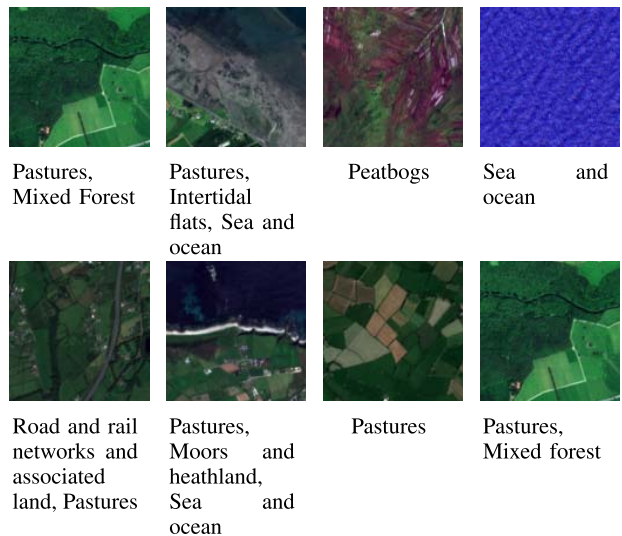

Fig. 4. Example of images with their multi-labels from the IR-BigEarthNet dataset.

Image labels (from figure):
- Pastures, Mixed Forest
- Pastures, Intertidal flats, Sea and ocean
- Peatbogs
- Sea and ocean
- Road and rail networks and associated land, Pastures
- Pastures, Moors and heathland, Sea and ocean
- Pastures
- Pastures, Mixed forest

TABLE I

Number of Samples Per Class in Training, Validation, and Test Sets of the IR-BigEarthNet Dataset

| Class | Train | Val | Test |
|---|---|---|---|
| Urban fabric | 818 | 369 | 371 |
| Arable lands | 2860 | 1404 | 1397 |
| Pastures | 5723 | 2725 | 2739 |
| Complex cultivation patterns | 510 | 240 | 226 |
| Land principally occupied by agriculture, with significant areas of natural vegetation | 896 | 420 | 414 |
| Broad-leaved forest | 410 | 205 | 199 |
| Coniferous forest | 1156 | 545 | 524 |
| Mixed forest | 588 | 273 | 296 |
| Moors, heathland and sclerophyllous vegetation | 467 | 217 | 219 |
| Transitional woodland, shrub | 708 | 345 | 369 |
| Inland wetlands | 811 | 400 | 415 |
| Marine waters | 2087 | 894 | 900 |

Seattle, Tampa, Tucson, and Ventura [56]. The size of each image is $256 \times 256$ pixels and it has a spatial resolution of 0.3 m. In the experiments, we used the multi-label annotations of UCMerced images that were obtained based on visual inspection [57]. The total number of class labels is 17, while the number of labels associated with each image varies between 1 and 7. The number of training, validation, and test samples in the multi-label UCMerced dataset is given in Table II. Fig. 5 shows an example of images together with their associated multi-labels from the UCMerced dataset.

### B. Experimental Setup

In the experiments, we used ResNet [58] as a backbone for the proposed RCML. Among different versions of ResNet, each of which includes different numbers of layers, ResNet50 with 50 layers was chosen. Moreover, we used the improved residual units [59] that enhance the network's generalization performance and make training easier by introducing additional nonlinearities. We used the stochastic gradient descent (SGD) optimizer with an initial learning rate of $10^{-3}$. We used a learning rate scheduler with an exponential decay rate of 0.9.

TABLE II

Number of Samples Per Class in Training, Validation, and Test Sets of the Multi-Label UCMerced Dataset

| Class | Train | Val | Test |
|---|---|---|---|
| Airplane | 70 | 15 | 15 |
| Bare Soil | 506 | 103 | 109 |
| Buildings | 482 | 102 | 107 |
| Cars | 631 | 125 | 130 |
| Chaparral | 77 | 21 | 17 |
| Court | 73 | 16 | 16 |
| Dock | 70 | 15 | 15 |
| Field | 73 | 15 | 15 |
| Grass | 683 | 145 | 147 |
| Mobile home | 70 | 17 | 15 |
| Pavement | 917 | 189 | 194 |
| Sand | 210 | 43 | 41 |
| Sea | 70 | 15 | 15 |
| Ship | 70 | 16 | 16 |
| Tanks | 70 | 15 | 15 |
| Trees | 702 | 155 | 152 |
| Water | 142 | 31 | 30 |

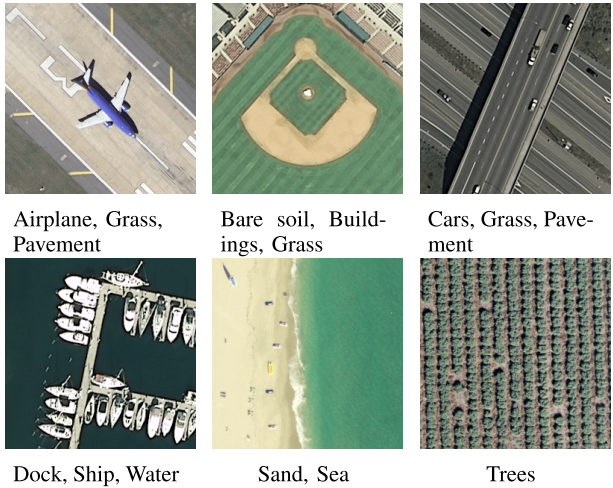

Airplane, Grass, Pavement    Bare soil, Buildings, Grass    Cars, Grass, Pavement

Dock, Ship, Water    Sand, Sea    Trees

Fig. 5. Example of images with their multi-labels from the multi-label UCMerced dataset.

The batch sizes for IR-BigEarthNet and multi-label UCMerced were set to 256 and 64, respectively. We report the results of training obtained after 100 epochs. The model training and further experiments were conducted on a Tesla V100 machine with 32-GB RAM.

To identify the potential noisy samples, a ranking loss for every sample was calculated using the group lasso module. The group lasso has two parameters $\alpha$ and $\beta$ that control the effects of two different noise types on the calculation of the ranking loss. We define $\beta = 1 - \alpha$. A group lasso module with a higher $\alpha$ concentrates more on finding the missing class labels, whereas a higher $\beta$ gives a higher weight to detect the wrong class labels. Since we do not consider any prior assumption, the values of $\alpha$ and $\beta$ were set to 0.2 and 0.8, respectively, based on a grid search strategy.

Within the swap module of the proposed RCML method, the networks exchange the calculated ranking information with each other. According to the ranking information and the swap rate $\gamma$, each network chooses a certain amount of samples associated with the smaller ranking loss values identified from the other network to update its weights. For each mini-batch, the identified samples associated with the highest ranking loss values are excluded from the final loss calculation. The goal of this process is to reduce the effect of noisy samples from backpropagation. Finding an appropriate value for the swap rate $\gamma$ is not easy, since it is highly related to the noise rate in the training set [40]. Therefore, the value of swap rate is defined as $\gamma = 1 - \hat{nr}$, where $\hat{nr}$ is the approximated noise rate estimated using a cross-validation approach.

Fig. 6. Considered label noise injection approach: RNS with a sampling rate 0.5 and a class rate of 0.5. The colored cells represent the introduced artificial label noise. Cells in blue represent "0" value in ground-truth labels that are flipped to "1," while those in red represent "1" value in ground-truth labels that are flipped to "0."

To assess the effectiveness of the proposed method, we added synthetic noise to the labels of the IR-BigEarthNet and multi-label UCMerced datasets. To ensure that both types of label noise (missing label and wrong label) are introduced to the multi-label training set, we designed a label noise injection approach called random noise per sample (RNS). RNS chooses labels randomly from the training set by a predefined percentage called the sampling rate. Afterward, to apply noise, RNS randomly flips a certain number of selected labels. The number of flipped labels driven by a parameter is called the class rate. An illustration of the considered label noise injection approach is shown in Fig. 6. The figure shows an example scenario for RNS with a sampling rate of 0.5 and a class rate of 0.5. Three samples out of six are selected, and label noise is randomly applied to half of their labels. In our experiments, the ratio of the introduced noise was varied from 0% to 50%.

The results of the experiments were provided in terms of two main performance metrics averaged over three runs for each experiments: 1) mean average precision (mAP) with the micro and macro averaging strategy and 2) $F_1$ score with the micro averaging strategy. For an explanation of these metrics, the reader is referred to [6]. Since the proposed RCML includes two networks that run simultaneously, the network with the best mAP micro validation score has been chosen for evaluation. To further study the performance and the behavior of the proposed RCML, the class-based mAP scores are reported in comparative plots. We compared our proposed method with four baseline methods: 1) FL [35]; 2) SAT; 3) ELR [28]; and 4) standard binary cross-entropy (BCE). To have a fair comparison for all the methods, we select the same architecture (ResNet50) pretrained on ImageNet under the same hyperparameters.

## V. Experimental Results

### A. Sensitivity Analysis

This section presents the sensitivity analysis for the proposed RCML method under different values of the

TABLE III

Sensitivity Analysis: The mAP Macro Scores (%) Obtained Under 10% and 40% Injected Noise Rates by Varying the Values of $\alpha$ and $\beta$ of RCML for the Multi-Label UCMerced Dataset

| injected noise rate | 10% | | | 40% | | |
|---|---|---|---|---|---|---|
| $\alpha$ ($\beta$=1-$\alpha$) | 0.2 | 0.5 | 0.8 | 0.2 | 0.5 | 0.8 |
| mAP macro (%) | **89.2** | 88.7 | 89.0 | **83.3** | 80.7 | 81.4 |

TABLE IV

Sensitivity Analysis: The mAP Macro Scores (%) Obtained Under 40% Injected Noise Rate by Varying the Values of $\lambda_2$ and $\lambda_3$ of RCML for the Multi-Label UCMerced Dataset

| $\lambda_2$ | $\lambda_3$ | mAP macro (%) |
|---|---|---|
| 1.0 | 1.0 | 81.6 |
| **1.0** | **0.1** | **83.3** |
| 0.1 | 1.0 | 79.5 |
| 0.5 | 0.5 | 80.7 |
| 0.1 | 0.1 | 80.3 |
| 0.01 | 0.01 | 79.6. |

TABLE V

Ablation Study: The mAP Macro Scores (%) Obtained Using Different Modules of the Proposed RCML and the CCML Methods Under 10% and 40% Injected Noise Rates on the Multi-Label UCMerced Dataset

| Injected noise rate | 10% | 40% |
|---|---|---|
| RCML all modules | **89.2%** | **83.3%** |
| RCML without MMD | 89.1% | 82.6% |
| RCML without group lasso | 89.2% | 77.0% |
| RCML without swap | 89.0% | 81.4% |
| Standard CCML | 82.1% | 81.2% |

hyperparameters. In the group lasso module, the two hyperparameters $\alpha$ and $\beta$ are the weights for balancing the missing and wrong class label ranking losses, respectively. For the sensitivity analysis, we choose $\alpha$ from a set of fixed values $\alpha = [0.2, 0.5, 0.8]$, while $\beta = 1 - \alpha$. We observe that $\alpha$ and $\beta$ are sensitive to: 1) the distribution of absence and presence classes in the label set (e.g., in our case, the majority of the class label for a given sample is absent) and 2) the type of multi-label noise in the training set (e.g., missing label or wrong class label). In the case of the availability of prior knowledge about the type of noise or class label distribution, we can use this knowledge to adjust $\alpha$ and $\beta$. In detail, a group lasso module with a higher value for $\alpha$ weights more on finding the missing class labels, whereas a higher $\beta$ gives more weight to identify the wrong class labels. Table III shows the sensitivity analysis results for $\alpha$ and $\beta$ for low (10%) and high (40%) injected noise rates. From the table, one can observe that the wrong class label is more harmful than a missing class label since increasing $\beta$ (the importance of noise in wrong class labels) improves the performance of the model. Thus, we set $\alpha$ to 0.2 and $\beta$ to 0.8. Also, by analyzing the results in Table III, one can see that the proposed RCML is robust to the variations in $\alpha$ and $\beta$ for two different noise rates.

The effect of different values for the hyperparameters $\lambda_2$ and $\lambda_3$ can be found in Table IV. The table shows that when $\lambda_2$ and $\lambda_3$ are weighted equally, decreasing the value of these hyperparameters affects the performance of the proposed RCML negatively. As an example, when the value of $\lambda_2$ and $\lambda_3$ is set to 0.01, the mAP macro score drops by 2%. Furthermore, choosing a higher value for $\lambda_2$ compared with $\lambda_3$ outperforms the case where the value of $\lambda_3$ is higher than $\lambda_2$. This shows that learning diverse features has a positive effect on the performance of the RCML. It is also observed that weighting both $\lambda_2$ and $\lambda_3$ equally with higher values does not increase the performance. These results are also confirmed in other experiments obtained using the IR-BigEarthNet dataset (not reported for space constraints).

### B. Ablation Study

To analyze the influence of each module, we designed different configurations by excluding individual modules from the proposed RCML. Furthermore, we compared RCML with the standard consensual collaborative multi-label learning (CCML) [22]. The standard CCML includes all the modules of RCML and also a flipping module that flips the noisy labels identified by both the networks during the final training epochs. In detail, we compared RCML with the standard CCML and three different configurations of RCML when we exclude: 1) the MMD module; 2) the group lasso module; and 3) the swap module. When removing the MMD module in the first configuration setup, there is no explicit objective to enforce two networks to learn different representations. In the second setup, to analyze the effect of the group lasso module, we use random sample selection instead of using the group lasso ranking information. When we remove the swap module, the two collaborative networks do not exchange the ranking information; instead, the weights of $f$ and $g$ are updated individually.

Table V shows the results of the ablation study in terms of mAP macro scores obtained when the injected noise rates are 10% and 40% for the multi-label UCMerced dataset. From the table, one can see that that the highest mAP is obtained when all the modules of RCML are included. We also observe that the MMD module has no significant influence on mAP. However, it is necessary to include MMD to ensure that two networks learn different representations and the obtained representations are not identical. Furthermore, the group lasso module has its biggest impact when the noise rate increases. When the injected noise rate is high (i.e., 40%), the ranking information obtained from group lasso is crucial to identify the noisy samples correctly, rather than a random exclusion. The other important module is the swap module, which aims to exchange the ranking information between two collaborative networks. When we exclude this module, each network is trained independently of the other network, which increases the risk of overfitting the noise since there is no consensual signal from the other network for correction. This becomes more severe when the noise rate is high, and the learning

TABLE VI

$F_1$, mAP Micro, and mAP Macro Scores Obtained by the Proposed RCML, FL, SAT, ELR, and BCE for the IR-BigEarthNet Dataset Under Different Noise Rates

| Noise | $F_1$ (%) | | | | | mAP micro (%) | | | | | mAP macro (%) | | | | |
|-------|------|------|------|------|------|------|------|------|------|------|------|------|------|------|------|
| Rate | FL | SAT | ELR | BCE | RCML | FL | SAT | ELR | BCE | RCML | FL | SAT | ELR | BCE | RCML |
| 0% | 64.0 | 66.0 | 66.2 | **72.5** | 72.3 | 78.5 | 77.6 | 80.6 | **81.0** | 80.9 | 45.5 | 43.8 | 47.7 | 49.6 | **49.8** |
| 10% | 62.4 | 65.0 | 65.2 | 71.0 | **72.3** | 77.8 | 76.6 | **80.1** | 79.2 | 79.7 | 45.1 | 43.0 | 47.0 | 47.5 | **47.8** |
| 20% | 55.9 | 62.4 | 62.3 | 67.9 | **71.7** | 77.3 | 74.4 | 79.0 | 77.4 | **79.2** | 43.8 | 41.0 | 45.5 | 45.5 | **47.2** |
| 30% | 48.5 | 56.2 | 57.0 | 61.4 | **70.8** | 76.6 | 71.2 | 77.6 | 75.4 | **79.0** | 42.7 | 38.3 | 43.6 | 42.8 | **46.6** |
| 40% | 39.0 | 48.3 | 48.4 | 56.3 | **70.3** | 75.6 | 69.0 | 76.0 | 76.5 | **79.0** | 40.7 | 36.0 | 41.4 | 41.7 | **46.5** |
| 50% | 30.7 | 34.0 | 35.1 | 37.5 | **67.8** | 70.1 | 69.3 | 72.8 | 73.7 | **78.5** | 35.3 | 34.2 | 37.3 | 37.9 | **45.3** |

process of the network(s) may distort with such a high noise rate in the multi-labels.

We also include the standard CCML to study the effect of improvements we made. The main drawback of the standard CCML is excluding a fixed portion of samples in each mini-batch (i.e., $\gamma = 25\%$) from backpropagation regardless of the injected noise rate. This can severely reduce the classification accuracy when the injected noise rate is low (e.g., less than 20%) since having a fixed swap rate $\gamma = 25\%$ excludes 25% of clean samples from backpropagation. We address this issue within the RCML by introducing an adaptive swap rate selection. RCML estimates the noise rate in the training set via cross-validation and sets the swap rate accordingly. As an example, when the noise rate is 10%, the standard CCML excluded 25% of the samples and obtained an mAP macro score of 82.1%, while RCML adaptively changes the swap rate to use all the clean samples and achieved 89.3%. These results are also confirmed in other experiments obtained using the IR-BigEarthNet dataset (not reported for space constraints).

### C. Comparison Among the Proposed RCML and Literature Methods

In this section, we compare the proposed RCML with the methods from the literature (FL [35], SAT [32], ELR [28], and BCE) for the IR-BigEarthNet and the multi-label UCMerced datasets. The results on IR-BigEarthNet are presented in Table VI. By analyzing the results in Table VI, one can see that the proposed method outperforms the comparison methods in all the evaluation metrics over most of the injected noise rates. Under a noise-free training set (0% injected noise rate), RCML provides comparable scores to the baseline BCE, while surpassing all the state-of-the-art noise robust methods. As an example, BCE provides an $F_1$ score of 83.4%, while RCML and SAT obtain $F_1$ scores of 83.3% and 80.2%, respectively. The results in Table VI also show that when the injected noise rate increases, the proposed RCML performs significantly better than the comparison methods. Especially under extreme noise rates such as 40% and 50%, RCML achieves in average 22% better $F_1$ scores and 6% better mAP macro scores than BCE. As an example under 50% injected noise, RCML provides mAP macro scores almost 30% and 15% higher than SAT and ELR, respectively. Furthermore, we observe that the performance of all the methods does not drop significantly when the injected noise rate is less than 20%. The reason is that the number of clean training samples for each class is still sufficient for the networks to learn and predict them correctly, despite the introduced label noise. The table shows that the mAP micro scores are higher than the mAP macro scores for all the methods. This reveals a class imbalance problem in the IR-BigEarthNet dataset. In this case, the classes with a small number of samples receive below-average scores because macro averaging computes the metric independently for class distributions and then takes the average, whereas micro averaging aggregates the contributions of each class.

For further analysis, we select six representative classes from IR-BigEarthNet and report their class-based mAP scores in Fig. 7. The class-based mAP scores in Fig. 7 reveal that the proposed RCML and the comparison methods learn the classes represented by sufficient training images better than the classes that do not include sufficient number of training images. The classes with a sufficient number of training images (e.g., "Marine waters") obtain the highest mAP scores under different noise rates. Even high noise rates, such as 40%, do not significantly reduce the mAP scores of these classes. As an example, for class "Marine waters" when the injected noise rate increases from 0% to 40%, RCML and ELR only dropped less than 1% in terms of mAP. The reason is that despite applying high noise rates, the classes with a high number of training images still have many images to learn from. On the other hand, the classes with insufficient training images (e.g., "Urban fabric," and "Moors, heathland and sclerophyllous vegetation") receive a low mAP score over different noise rates. As an example, for class "Urban fabric" when the injected noise rate increases from 0% to 40%, the mAP scores obtained by RCML and SAT dropped about 5% and 25%, respectively. The proposed RCML outperforms the comparison methods in the majority of classes, especially under high noise rates, while obtaining comparative or better results compared with the BCE in smaller noise rates (under 20%). For few classes (e.g., "Arable lands" and "Pasture"), the BCE and RCML performances are comparable over the low rates of the injected label noise. However, by increasing the noise rate, the RCML shows stability and obtains mAP scores higher than the other methods.

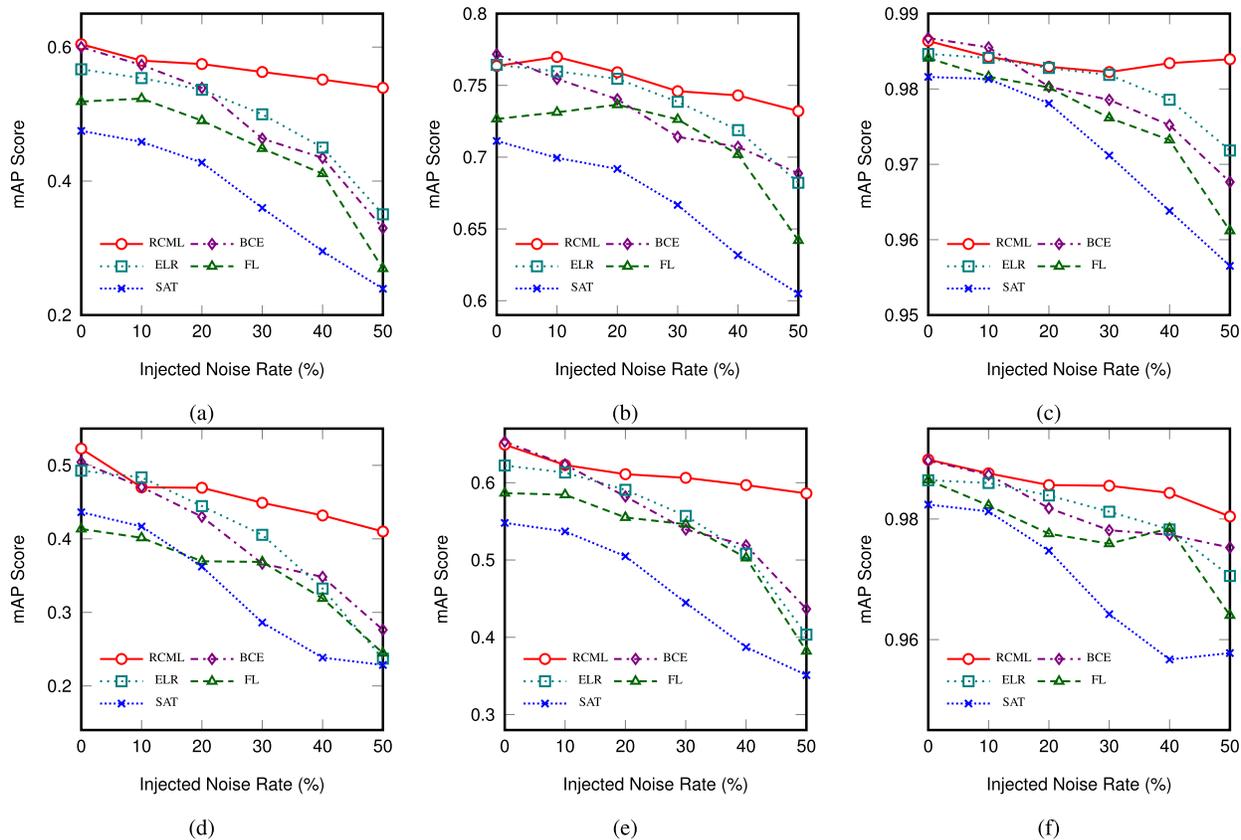

Fig. 7. Class-based comparison: mAP scores for different noise rates, obtained by the proposed RCML, BCE, ELR, FL, and SAT for six selected classes: (a) Urban fabric; (b) Arable lands; (c) Pasture; (d) Moors, heathland, and sclerophyllous vegetation; (e) Inland wetlands; (f) Marine waters in the IR-BigEarthNet dataset.

TABLE VII

$F_1$, mAP Micro, and mAP Macro Scores Obtained by the Proposed RCML, FL, SAT, ELR, and BCE in the Multi-Label UCMerced Dataset Under Different Noise Rates

| Noise | $F_1$ (%) | | | | | mAP micro (%) | | | | | mAP macro (%) | | | | |
|---|---|---|---|---|---|---|---|---|---|---|---|---|---|---|---|
| Rate | FL | SAT | ELR | BCE | RCML | FL | SAT | ELR | BCE | RCML | FL | SAT | ELR | BCE | RCML |
| 0% | 78.8 | 80.2 | 79.1 | **83.4** | 83.3 | 85.1 | 86.8 | 90.2 | 90.8 | **91.1** | 81.3 | 83.4 | 89.3 | **90.0** | 89.8 |
| 10% | 78.2 | 79.9 | 78.5 | **83.1** | 83.1 | 84.5 | 86.3 | 89.9 | 90.5 | **90.7** | 80.3 | 82.6 | 89.0 | **89.5** | 89.2 |
| 20% | 74.4 | 77.2 | 74.1 | 82.4 | **82.7** | 81.3 | 83.7 | 88.8 | 88.8 | **89.4** | 76.1 | 78.8 | 87.5 | 86.7 | **87.6** |
| 30% | 67.3 | 70.6 | 64.0 | 79.0 | **82.1** | 75.3 | 78.5 | 86.1 | 85.4 | **88.6** | 67.4 | 71.1 | 82.6 | 80.3 | **85.7** |
| 40% | 60.7 | 63.8 | 55.2 | 74.4 | **80.4** | 69.0 | 73.0 | 83.0 | 81.8 | **87.3** | 59.5 | 63.9 | 77.8 | 74.5 | **83.3** |
| 50% | 50.0 | 51.8 | 44.3 | 62.6 | **78.2** | 54.6 | 59.0 | 73.2 | 70.3 | **85.5** | 45.2 | 48.2 | 64.2 | 60.5 | **79.4** |

This demonstrates the robustness of RCML under different values of injected noise rates.

The comparison results for the multi-label UCMerced dataset are presented in Table VII. By analyzing the table, one can see that increasing the injected noise rates severely affects the learning processes in the multi-label UCMerced dataset. As shown in Table VII, label noise creates instability on the comparison methods. This is mainly due to the smaller training set of the multi-label UCMerced dataset. In general, the proposed RCML provides relatively stable performance and achieves a degree of robustness against different rates of the label noise. Although ELR and BCE achieve comparable mAP micro and mAP macro scores to the proposed RCML under lower noise rates, the RCML achieves considerably higher mAP and $F_1$ scores under 40% and 50% noise rates. As an example, under 10% injected noise rate BCE obtains an mAP macro score of 89.5% that is marginally better than RCML (0.3% higher). However, when the injected noise rate increases to 40%, the mAP macro score obtained by BCE drops to 74.5%, while the proposed RCML obtains an mAP macro score of 83.3%. We should note that ResNet50 is a powerful model that is capable of tolerating even a relatively high noise rate, but it is not stable under extreme noise rates. Using ResNet50 as a classifier within our proposed collaborative framework makes it more stable and robust against higher noise rates. A comparison between Tables VI and VII reveals that the multi-label UCMerced dataset achieves higher mAP and $F_1$ scores compared with IR-BigEarthNet under the same experimental setup. The reason for this is that the features in the images in the multi-label UCMerced dataset are more similar to the features that are learned in the ImageNet weights. This makes it easier for the classifiers to learn the underlying class distribution from the training set.

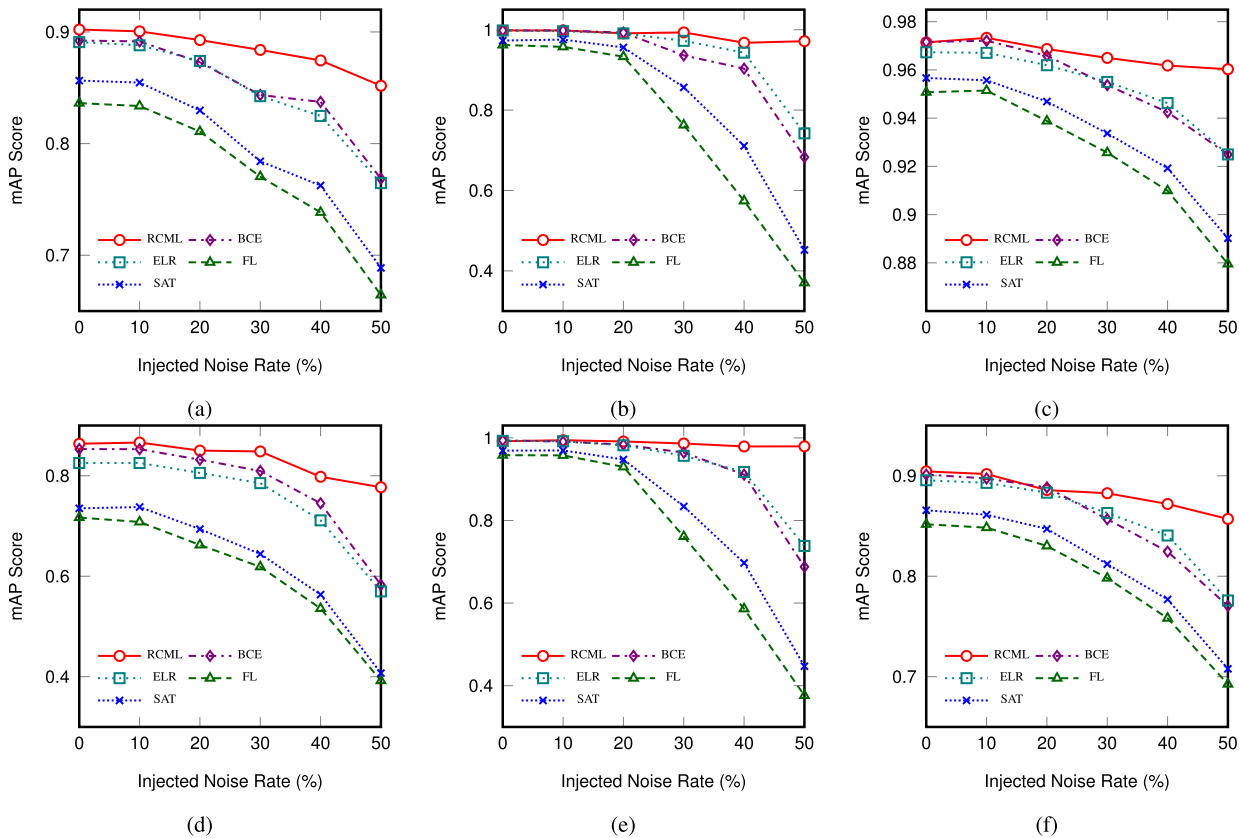

Fig. 8. Class-based comparison: Class-based mAP scores for different noise rates obtained by the proposed RCML, BCE, ELR, FL, and SAT for six selected classes: (a) Cars; (b) Dock; (c) Pavement; (d) Sand; (e) Ship; (f) Trees in the multi-label UCMerced dataset.

We analyze the class-based performances also for the multi-label UCMerced dataset. Fig. 8 shows mAP versus different injected noise rates for six representative classes from the multi-label UCMerced dataset. The performance of the comparison methods is not stable under different noise rates, as shown in Fig. 8. As an example, in the case of the "Dock" class, the mAP scores obtained by the proposed RCML are relatively stable (around 98%), while the performances of the comparison methods sharply drop when the noise rate exceeds 20% (e.g., the mAP scores of ELR and SAT are reduced more than 20% and 45%, respectively). Similar to the IR-BigEarthNet results (see Fig. 7), the mAP scores of the classes with a sufficient number of training samples are the highest compared with the other classes.

## VI. Conclusion and Discussion

In this article, we have proposed an RCML method to overcome the negative effects of multi-label noise in the context of scene classification of RS images. The proposed method includes three main modules: 1) discrepancy module; 2) group lasso module; and 3) swap module. The discrepancy module forces the two networks to learn diverse features while obtaining consistent predictions. The group lasso module detects the noisy labels assigned to the training images. Using the group lasso module, the RCML method identifies the different types of multi-label noises that are associated with missing and wrong class label annotations. Finally, the swap module exchanges the ranking information between two networks and excludes the identified samples associated with noisy labels from backpropagation dynamically. To the best of our knowledge, RCML is the first method that simultaneously tackles the negative effects of the two types of multi-label noise without making any prior assumption.

The performance of the proposed method under different noise rates was evaluated on two publicly available multi-label benchmark RS image archives that are the IR-BigEarthNet and the multi-label UCMerced archives. The experimental results confirm the effectiveness of the proposed method in specific settings where deterministic label noise is introduced to the multi-label training sets. Furthermore, RCML results in higher accuracy with respect to other state-of-the-art methods under the presence of high multi-label noise rates such as 30% and more.

We would like to point out that developing efficient techniques for handling label noise in multi-label training sets is becoming more and more important. On one side, due to the increased volume of RS image archives, manual large-scale image labeling is time-demanding and costly (and thus not fully feasible). On the other side, making use of zero-cost labeling by the use of available thematic products can introduce label noise in the training sets. In this context, the proposed RCML is very promising as it allows to identify the potential multi-label noise within the training sets without considering any prior assumption. We observe that RCML has identified more than 65% (in average over all noise rates) noisy samples correctly, and excluding the identified noisy samples from backpropagation could significantly improve the performance of MLC. Furthermore, the proposed method is intrinsically classifier-independent. In our experimental setup, RCML is implemented in the context of CNNs (because of their efficiency for RS image classification); however, RCML can be adapted easily for any other classifier or network architecture. In the case of using a different network architecture, the layers where the discrepancy module will be inserted should be carefully selected. To this end, one can apply a hyperparameter tuning for the considered network architecture to find the most appropriate layers for inserting the discrepancy module.

As a final remark, it is worth noting that the proposed RCML provides more accurate results when the approximated noise rate is close to the injected noise rate in the training set. The noise rate approximation is essential since it determines the swap rate value. As a future work, we plan to extend the proposed RCML to consider the noise rate as an extra parameter and learn it during end-to-end training.

## Acknowledgment

The authors would like to thank Tristan Kreuziger for the valuable discussions during the initial phase of this work.

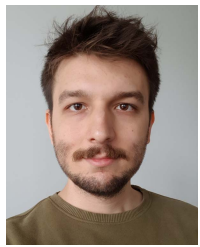

**Ahmet Kerem Aksoy** (Student Member, IEEE) received the B.S. degree in computer science from the Technical University of Berlin, Berlin, Germany, in 2020, where he is currently pursuing the M.S. degree in computer science.

He joined the Remote Sensing and Image Analysis Group there as a Research Assistant during his graduate study. His research interests include computer vision, machine learning, and remote sensing.

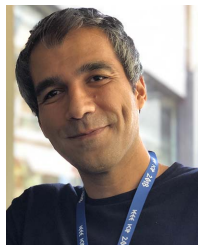

**Mahdyar Ravanbakhsh** (Member, IEEE) received the Ph.D. degree from the University of Genoa, Genoa, Italy, in 2019.

He was a Post-Doctoral Research Fellow with the University of Genoa and a Visiting Researcher with the Deep Relational Learning Group, University of Trento, Trento, Italy, in 2016. He was a Research Fellow with the Remote Sensing Image Analysis (RSiM), Department of Electrical Engineering and Computer Science, Technische Universität Berlin, Berlin, Germany, from 2019 to 2022. He is a Research Scientist with Zalando SE, Berlin. His research lies at the intersection of machine learning and computer vision with an emphasis on deep learning with minimal supervision and/or limited data.

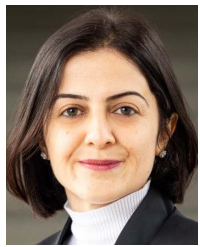

**Begüm Demir** (Senior Member, IEEE) received the B.S., M.Sc., and Ph.D. degrees in electronic and telecommunication engineering from Kocaeli University, Kocaeli, Turkey, in 2005, 2007, and 2010, respectively.

She is currently a Full Professor and the Founder Head of the Remote Sensing Image Analysis (RSiM) Group, Faculty of Electrical Engineering and Computer Science, TU Berlin, Berlin, Germany, and the Head of the Big Data Analytics for Earth Observation Research Group, Berlin Institute for the Foundations of Learning and Data (BIFOLD), Berlin. Her research activities lie at the intersection of machine learning, remote sensing, and signal processing. Specifically, she performs research in the field of processing and analysis of large-scale Earth observation data acquired by airborne and satellite-borne systems.

Dr. Demir is a Scientific Committee Member of several international conferences and workshops, such as Conference on Content-Based Multimedia Indexing, Conference on Big Data from Space, Living Planet Symposium, International Joint Urban Remote Sensing Event, SPIE International Conference on Signal and Image Processing for Remote Sensing, and Machine Learning for Earth Observation Workshop organized within the European Conference on Machine Learning (ECML)/Principles and Practice of Knowledge Discovery (PKDD). She was awarded the Prestigious "2018 Early Career Award" by the IEEE Geoscience and Remote Sensing Society for her research contributions in machine learning for information retrieval in remote sensing. In 2018, she received a Starting Grant from the European Research Council (ERC) for her project "BigEarth: Accurate and Scalable Processing of Big Data in Earth Observation." She is a Fellow of European Lab for Learning and Intelligent Systems (ELLIS). She is a Referee for several journals such as the PROCEEDINGS OF THE IEEE, the IEEE TRANSACTIONS ON GEOSCIENCE AND REMOTE SENSING, the IEEE GEOSCIENCE AND REMOTE SENSING LETTERS, the IEEE TRANSACTIONS ON IMAGE PROCESSING, *Pattern Recognition*, the IEEE TRANSACTIONS ON CIRCUITS AND SYSTEMS FOR VIDEO TECHNOLOGY, the IEEE JOURNAL OF SELECTED TOPICS IN SIGNAL PROCESSING, the *International Journal of Remote Sensing*, and several international conferences. She is currently an Associate Editor for the IEEE GEOSCIENCE AND REMOTE SENSING LETTERS, *MDPI Remote Sensing*, and the *International Journal of Remote Sensing*.